\documentstyle[epsfig]{qcdparis}
\begin{document}
\pagestyle{plain}
\title{
Two-component colour dipole emission
in the central region of onium-onium scattering
}

\author{
 R. Peschanski}

\affil{ Service de Physique Th\'eorique, CE-Saclay\\

F-91191 Gif-sur-Yvette Cedex, FRANCE}

\abstract{
The initial-state radiation of soft colour dipoles produced in the
central region of onium-onium scattering via single QCD
Pomeron exchange (BFKL) is calculated in the
framework of  Mueller's dipole approach. The resulting dipole production has
a two-component structure. One is constant with energy
while the  other grows
and possesses a power-law tail
at appreciably large transverse distances from the collision axis.
It may be related to the growth of the gluon distribution at small Bjorken-$x.$
}
\resume{
l' \' emission de dipoles de couleur dans la r\' egion centrale
d' une collision onium-onium par \' echange de la singularit\' e
BFKL de la QCD est calcul\' ee dans la r\' ecente  approche des dipoles
due \` a Mueller.
Une double composante de la section efficace est obtenue par le
calcul; L' une est constante en \' energie
alors que l' autre augmente et poss\` ede une distribution en loi de puissance
dans la coordon\' ee transverse \` a l' axe de la r\' eaction. Cette composante
est probablement associ\' ee \` a la croissance de la distribution des gluons
\` a petite valeur du param\` etre $x$ de Bjorken.
}
\twocolumn[\maketitle]
\fnm{7}{Talk given in the
Hadronic Final States session
at the Workshop on Deep Inelastic scattering and QCD,
Paris, April 1995}
\section{Introduction}
High energy onium-onium scattering is a simple process which can be used to
study the physics of the perturbative QCD Pomeron, the so-called BFKL
singularity~\cite{rose1}.  Recently a quantitative picture
in which a high-$ Q^2 $ $ q\bar q $ (or {\it onium}) state
looks like a
collection of colour dipoles of various sizes
has been developped
by  Mueller~\cite{rose2,rose3,rose4}.
 The QCD Pomeron
elastic amplitude is recovered in this dipole picture provided the onium-onium
elastic scattering comes from a dipole in one onium state scattering off a
dipole in the other onium state by means of two-gluon exchange
{}~\cite{rose2,rose3,rose4,rose5}.
 In this paper we elaborate on some consequences
of the dipole approach for the initial-state radiation associated with a
single Pomeron exchange.

Our starting point is the observation that the onium-onium
scattering process is accompanied by a radiation due to colour dipoles
which -- while present in the initial state -- are released during the
collision. In this contribution we present an explicit computation of
this dipole emission process
following  Mueller's approach, more
specifically that of Ref.~\cite{rose4}. It follows essentially
from a recent paper~\cite{rose6}  by Andrzej Bialas and myself.
As an addendum to the published paper we
present an interesting  further consequence of this calculation, namely the
{\it
two-component}
structure of the predicted dipole radiation.

We intend to estimate the emission of
colour dipoles from the initial-state. To
this end, we first write the formula for the inclusive cross-section
for emission of a dipole in the central region from one of the colliding onia.
It reads:

$$ \eqalignno{& x\
{ {\rm d} \sigma \over {\rm d} x {\rm d} r^2}
 = 4\pi
\alpha^ 2 \int^{ \infty}_ 0{ {\rm d} l \over l^3}\ { {\rm d} x_1 \over x_1}
\ { {\rm d}x_2 \over x_2}  \left[1-J_0 \left(l x_1
\right) \right]
&  \cr  &
\times  \left[1-J_0 \left(l x_2 \right) \right]
\int^{ }_{ }
{\rm d}^2b\
{\rm d}^2b_1\
{\rm d}^2b_2\
 \delta^2(b-b_1-b_2)
&  \cr  &
\times \ n^{(1)}
\left(x_{ 01},b_1, x_1 ;{Y\over 2}\right)
n^{(2)}
\left(x_{02}, b_2, x_2; r,x;{Y\over 2} \right) & (1) \cr} $$
where $ n^{(1)} $ is the {\it single} dipole density in the onium of size $
x_{ 01} $ and $ n^{(2)} $
is the {\it double} dipole density in the onium of size $ x_{01}, $
both dipoles being present  in the central region of rapidity $Y/2.$

This formula -which is an extension of Mueller's
formula for the total onium-onium cross-section (see Eq.(3) of
Ref.~\cite{rose4} -
expresses simply
the fact that the number of emitted dipoles at rapidity $Y/2$
is just the number of dipoles present at that same rapidity
in the initial state
{\it whenever} the interaction took place.
The important feature of Eq.(1) is that it contains the double-dipole density
$ n^{(2)} $ at central rapidity coming from one of the initial onia (since one
of the two dipoles
is involved in the interaction mechanism). Thus, it is sensitive to
 correlations between dipoles in the same onium state. As
we shall see, this has non-trivial consequences. In fact, already
 at this stage, one may expect that these correlations should be
rather strong because, as shown in Refs.~\cite{rose2,rose3,rose4}, the colour
dipoles
which contribute to the onium wave function are formed in a cascade
process and thus cannot be independent. Furthermore, since this cascade is
scale-invariant one also expects scale invariance in the dipole-dipole
correlations to be valid, which -in turn- is likely to be a reflection of the
inner conformal invariance, known to be rooted in the formalism
of the BFKL Pomeron~\cite{rose7}.

\section{The dipole emission cross-section}
\subsection{The Mellin Transform of the cross-section}
The appropriate expression for $ n^{(1)}, $ was given in Ref.~\cite{rose4}:
$$ \eqalignno{ n^{(1)} \left(x_{ 01},b,
x;Y/2 \right) & \equiv  {x_{ 01} \over 4x
\left\vert b \right\vert^ 2}\ {\rm ln}{ \left\vert
b\right\vert^ 2 \over x_{ 01}  x}
\ {\rm
exp}  \left(-a\ {\rm ln}^2{ \left\vert b \right\vert^ 2 \over
x_{ 01} x}\right) &  \cr  &  \times  {\rm exp} \left[
\left(\alpha_{ p}-1 \right)Y/2 \right] \left({4a \over \pi} \right)^{3/2}\ ,
& (2) \cr} $$
where $$
 a \equiv a(Y) = \left[7\alpha \ N_c\zeta( 3)Y/\pi \right]^{-1}
\approx   \left[ 3 \left(\alpha_ p-1 \right)Y\right]^{-1}.\eqno
 (3) $$

The needed explicit expression for $ n^{(2)} $ has been derived in
our paper~\cite{rose6} extending the formalism of
Ref.~\cite{rose4}.
It makes use of an inverse Mellin transform, that is:
$$ \eqalignno{
 n^{(2)} \left(x_{02}, b_2,x_2; r,x;Y \right)& = \int^{c+i\infty }_
{ c-i\infty}{
{\rm d} \gamma \over 2i\pi}  \ x_{02}^{\gamma}\ \times
&  \cr  & \times \tilde n^{(2)} \left(\gamma , b_2, x_2;
r,x;Y \right)
& (4) \cr
{ {\rm d} \sigma \over {\rm d} x {\rm d} r^2} (x_{02}) & =
\int^{c+i\infty }_
{ c-i\infty}{
{\rm d} \gamma \over 2i\pi}  \ x_{02}^{\gamma}\
{ {\rm d}\tilde  \sigma \over {\rm d} x {\rm d} r^2} (\gamma)
 &  \cr}
 $$
with the real constant $c < 2$ for convergence condition.

Using the hypergeometric function
$$ \ _2F_1 \left[y^2 \right] \equiv  \ _2F_1 \left({\gamma \over 2},
{\gamma \over 2};1;y^2 \right),$$
one obtains as a final result:
$$ \eqalignno{ { {\rm d}\tilde \sigma \over {\rm
d} x {\rm d} r^2} & =  \sigma_{ {\rm tot}} \
2\sqrt{ 2}\
 {\alpha N_c \over \pi}
\left({2a \over \pi}
\right)^{3/2}
\int_0^1{{\rm d} y\ _2F_1\left[y^2\right]}
& \cr
 & \times \ {r^{-\gamma} \over x}\ {{\rm
exp} \left[
\left({\alpha N_c \over \pi} \chi (\gamma) - (\alpha_ p-1) \right)\ Y/2
\right] \over
2\ \left(\alpha_ p-1 \right)\
-\ {\alpha N_c \over \pi} \chi (\gamma)}
 & (5) \cr  &  \times
\ { {\rm ln}\left( {r \over x}\right)
\ {\rm exp}\left[-a {\rm ln}^2 \left({r
\over x} \right)\right]
 \over 2 +
2 a {\rm ln} \left({r
\over x} \right)-\ \gamma }
.&  \cr} $$
where, by definition,
$$ \chi \left(\gamma \right) \equiv  2\psi( 1) - \psi \left(1-\gamma/2
\right)-\psi \left(\gamma/2 \right)
, \eqno (6) $$
$ \psi( \gamma)  \equiv  {{\rm d}
\over {\rm d}
\gamma}
 {\rm ln} \ \Gamma
,\ \chi (1)\ {\alpha N_c \over \pi}\equiv \alpha_p -1
= \  4 {\rm ln}  2\ {\alpha N_c \over \pi}
. $
\subsection{
The two components of the cross-section}

The inverse Mellin transform of (5) has to be
determined in order
to obtain the expression for
the emission cross-section
${ {\rm d}\sigma \over {\rm d} x
{\rm d} r^2}.$
In performing it, one has to take into account the different
singularities in the $\gamma -$variable present in expression (5). There are
of two kinds, namely the
poles in the denominators, and the well-known saddle-point in the function
$\chi(\gamma)$ at $\gamma = 1.$
These contributions lead to a
two-component structure of the dipole radiation in the central region.
\subsubsection{ Component I : the saddle-point }

The saddle-point contribution comes from the second row of
the expression
(5). One gets:
$$
\left({ {\rm d} \sigma \over {\rm d} x
{\rm d} r^2}\right)_I
 = \ \sigma_{ {\rm tot}}\
{1 \over xr^2}\
 \left({r \over
x} \right) \varphi _I\left(
{x  \over r},
{ x_{10} \over r}
\right)\ , \eqno (7) $$
where~\cite{rose4} $\sigma_{{\rm tot}} = 2 \pi
\ x_{10}x_{20}\ \alpha ^2
\ {\rm e}^{(\alpha_p-1)\ Y}\ \left({2 a \over \pi}\right)^{1/2},
$
$$ \varphi _I \approx {\cal C}st. \
\left({2a \over \pi}
\right)^2
\ {\rm ln}\ {r \over x}\ {\rm e}^{ -a\
\left({\rm ln}^2 {r \over x}
+ \ {\rm ln}^2 {r \over x_{10}} \right)}
\ . \eqno (8) $$

\subsubsection{
 Component II : the pole }

One has to take into account the poles present in the denominators of
expression (5).
 In fact the one which is
less but closest to the $2  \pm i\infty$ line in the complex $\gamma$-plane
will dominate.
One obtains:
$${\alpha N_c \over \pi}\ \chi
\left(\gamma^{\ast}\right) =
2 (\alpha_p \ - \ 1 ) \equiv
\ {2\ \alpha N_c \over \pi}\ \chi(1),
$$
Looking for a solution of the form $\gamma^{\ast} = 2 - \gamma_M$
and using the explicit form of the kernel $\chi(\gamma)$ leads to
a value $\gamma _M \approx .37.$
 Notice that the other denominator in (5)
corresponds to a pole outside the integration contour (at $\gamma > 2).$

 One finally gets:
$$
\left({ {\rm d} \sigma \over {\rm d} x
{\rm d} r^2}\right)_{II}
 = \ \sigma_{ {\rm tot}}\
{1 \over xr^2}\
{\rm e}^{\left(
\left(\alpha_ p-1 \right){Y \over 2} \right)}
\ {x_{10} \over x} \left({r \over
x_{10}} \right)^{\gamma _M} \varphi _{II}
, \eqno (9) $$
$$ \varphi _{II}\approx {\cal C}st.
\left({2a \over \pi}
\right)^{{3 \over 2}}
\ {\rm ln}\ {r^2 \over xx_{20}}\ {\rm e}^{ -a
{\rm ln}^2 \left({r^2 \over xx_{20}}
 \right)}
. \eqno(10) $$

\section{Conclusion}
The interpretation of the two components (I and II)
of the inclusive cross-section in terms of
conventional gluon diagrams deserves more study and
their specific properties bring some unexpected
features. Let us briefly outline some intersting
aspects.

The first component is approximately independant of the energy
apart logarithmic factors. It is tempting to associate it
with a similar contribution found in the derivation of the
Pomeron-Pomeron contribution for gluons in the BFKL formalism~\cite{
rose8}.
By contrast, the second component has a rather strong energy dependence.
In fact, it is related to the large rapidity difference between the initial
dipole $x_{20}$ and the one emitted in the very central region near the
rapidity $Y/2.$
 It means
a very small value
of $x_{BJ} \approx e^{- Y/2}$
and consequently, the inclusive cross-section is
sensitive to the singular behaviour of the gluon distribution
at small $x_{BJ}$ related to the BFKL singularity.  The properties of this
component have been
examined in detail in our paper~\cite{rose6}, showing in particular
the interesting feature of a power-law tail in the transverse distance with
respect to the collision axis.
\begin{center}
{\large\bf Aknowledgements}
\end{center}
All results come from a fruitful and friendly collaboration
with Andrzej Bialas.

\vspace{2cm}

\end{document}